\documentclass[a4paper,twoside]{article}

\usepackage{epsfig}
\usepackage{subcaption}
\usepackage{calc}
\usepackage{amssymb}
\usepackage{amstext}
\usepackage{amsmath}
\usepackage{amsthm}
\usepackage{multicol}
\usepackage{pslatex}
\usepackage{apalike}
\usepackage{alltt}
\usepackage[utf8]{inputenc}
\usepackage[hidelinks]{hyperref}
\usepackage{SCITEPRESS}

\begin{document}

\title{Bridging the Technology Gap Between Industry and Semantic Web:\\Generating Databases and Server Code From RDF}

\author{\authorname{Markus Schröder\sup{1,2}, Michael Schulze\sup{1,2}, Christian Jilek\sup{1,2} and Andreas Dengel\sup{1,2}}
	\affiliation{\sup{1}Smart Data \& Knowledge Services Dept., DFKI GmbH, Kaiserslautern, Germany}
	\affiliation{\sup{2}Computer Science Dept., TU Kaiserslautern, Germany}
	\email{\{markus.schroeder, michael.schulze, christian.jilek, andreas.dengel\}@dfki.de}
}


\abstract{
Despite great advances in the area of Semantic Web, industry rather seldom adopts Semantic Web technologies and their storage and query concepts.
Instead, relational databases (RDB) are often deployed to store business-critical data, which are accessed via REST interfaces.
Yet, some enterprises would greatly benefit from Semantic Web related datasets which are usually represented with the Resource Description Framework (RDF).
To bridge this technology gap, we propose a fully automatic approach that generates suitable RDB models with REST APIs to access them.
In our evaluation, generated databases from different RDF datasets are examined and compared.
Our findings show that the databases sufficiently reflect their counterparts while the API is able to reproduce rather simple SPARQL queries.
Potentials for improvements are identified, for example, the reduction of data redundancies in generated databases.
}

\onecolumn \maketitle \normalsize \setcounter{footnote}{0} \vfill

\section{\uppercase{Introduction}}
\label{sec:introduction}

\noindent The Resource Description Framework (RDF) \cite{RDF1.1primer} is a well-established data model in the Semantic Web community.
It is used to express facts about resources identified with uniform resource identifiers (URIs) \cite{RFC2396URI} in the form of statements (subject, predicate and object).
Ontologies \cite{Gruber1993TranslationApproachtoPortableOntologySpec} are used to model domains and to share formally specified conceptualizations which can be expressed by using RDF Schema (RDFS) \cite{RDFSchema}.
RDF-based data is typically stored in triplestores and queried with the SPARQL Protocol and RDF Query Language (SPARQL) \cite{SPARQL1.1overview}.

In our experience, beyond the semantic web and especially in industry, such technologies are rather seldom used. 
A case study for the manufacturing industry also points out this observation \cite{DBLP:journals/dke/FeilmayrW16}.
Instead, relational databases (RDBs) are commonly used to store important and system critical data -- information systems which are well-researched over 50 years. By implementing Application Programming Interfaces (APIs), a controlled access on these datasets with Create, Read, Update and Delete (CRUD) operations are provided for system developers.
APIs often conform to the Representational State Transfer (REST) software architecture, utilize the Hypertext Transfer Protocol (HTTP) and exchange data in the JavaScript Object Notation (JSON) format.

By examining both sides, we observe distinct solutions regarding storage and query approaches.
\autoref{tbl:comparison} summarizes and compares these findings:
while the Semantic Web encourages the use of triplestores loaded with ontologies and RDF statements, industry prefers databases with defined schemata and stored records.
Graph-oriented SPARQL queries and their special result set responses are opposed to document-oriented REST APIs returning JSON documents.
\begin{table}
	\centering
	\caption{
		Comparison of storage and query concepts between Semantic Web and Industry.
	}
	\resizebox{\columnwidth}{!}{\begin{tabular}{|l||c|c|}
		\hline
		Concept & Semantic Web & Industry \\
		\hline
		\hline
		Storage & Triplestore & Database \\
		\hline
		Domain Modeling & Ontology & Database Schema \\
		\hline
		Data Modeling & RDF Statements & Database Records \\
		\hline
		Identification & URIs & Numeric IDs \\
		\hline
		Query Interface & SPARQL & REST API \\
		\hline
		Exchange Format & SPARQL Result Set / RDF & JSON \\
		\hline
	\end{tabular}
	}
	\label{tbl:comparison}
\end{table}

Yet, some enterprises would greatly benefit from Semantic Web technologies. 
This also includes related datasets and the way knowledge is modeled.
We see a trend that more and more publicly available datasets are modeled and/or published in the RDF format, such as datasets in the Linked Open Data (LOD) cloud\footnote{\url{https://lod-cloud.net/}}, DBpedia \cite{Bizer2009DBpedia} and Wikidata \cite{Vrandecic2014Wikidata}.
Moreover, it has been become popular in data integration services and especially in knowledge services to construct and maintain knowledge graphs \cite{hogan2020knowledge} for selected use cases, for instance, when building a corporate memory \cite{Maus2013}.
To embed these new datasets and technologies in workflows and processes, corporations would have to spend considerable efforts.
In order to keep a company's overhead to a minimum, we suggest transforming RDF-based datasets back to storage systems they are more familiar with, namely RDBs.
For integration purpose, we further recommend that enterprises implement CRUD REST APIs to provide access and manipulation layers for their developers.
Since such conversions and implementations are quite tedious and costly when executed manually, a fully automatic way to generate the envisioned assets would be helpful.

Since related work did not appropriately address this particular use case, in this paper, we provide a solution to this challenge.
A generation procedure is described that accepts an arbitrary RDF(S) dataset and generates an RDB with a CRUD REST API to access and modify it.
Doing this, raises the following research questions which are addressed in our experiments:
\begin{enumerate}
	\item How well do the generated RDBs reflect their RDF dataset counterparts?
		By using various RDF datasets, we check if any critical data is missing in the databases and how they are structured.
	\item How well can the generated CRUD REST API reproduce queries that would have been performed with SPARQL?
		To answer this, we try to query same information with our API compared with given SPARQL queries. 
	\item What limitations do the generated databases and interfaces have?
		In our experiments, we reveal and discuss shortcomings of our approach.
\end{enumerate}
For future research, the source code of our algorithm and the evaluation material is publicly available at GitHub\footnote{\url{https://github.com/mschroeder-github/rdf-to-rdb-rest-api}}.

This paper is structured as follows:
in the next section (Sec. \ref{sec:relwork}) we investigate procedures and tools in literature that also transform RDF to RDB.
This is followed by our own approach in Section \ref{sec:approach}.
Section \ref{sec:eval} presents the evaluation of our method and answers the stated research questions.
We close the paper with a conclusion and an outlook in Section \ref{sec:concl}.

\section{\uppercase{Related Work}}
\label{sec:relwork}

\noindent One can find a lot of papers in literature which are related to the conversion of RDB (and similar formats) to RDF.
However, only few works actually investigated the opposite direction (from RDF to RDB).

An early work \cite{teswanich2007transformation} transforms RDF documents to databases to apply Business Intelligence (BI) technologies. 
RDFS-related information is stored in meta tables, for example, in relations like \textit{class}, \textit{property} and \textit{sub\_class\_of}.
For each class and object property (regardless of its cardinality) a table is created. SQL statements demonstrate how a generated database can be queried.

Similarly, the R2D approach \cite{DBLP:conf/semco/RamanujamGKST09} generates databases from RDF data to reuse visualization tools.
The authors suggest several improvements regarding \cite{teswanich2007transformation}:
in contrast, their approach still works, albeit no ontological information is available in the input dataset.
Moreover, it considers the cardinality of properties to avoid the creation of tables, and it also handles blank nodes.

RDF2RDB\footnote{\url{https://github.com/michaelbrunnbauer/rdf2rdb}} is a Python based tool that converts a given RDF/XML document into a MySQL database.
The generation approach is also comparable with the one from \cite{teswanich2007transformation}, except it does not generate meta tables for RDF schema information.

A master thesis \cite{distr} investigated how RDF data can be converted to relational databases as well.
The proposed procedure is comparable with the previously mentioned works:
it scans the assertion box (A-Box) statements for predicates and instances to infer a database schema.
Different strategies are proposed how the arrangement of many-to-many tables could be achieved.

RETRO \cite{rachapalli2011retro} focuses on query translation with the same motivational arguments as we have about bridging the gap between Semantic Web and industry.
However, their approach does not physically transform RDF to an RDB.
Instead, they use a fixed schema mapping approach that virtually maps all predicates from the A-Box statements to relational tables having two columns, namely for subject and object.

Although R2D and RDF2RDB are comparable to our approach, we did not find any related work that also takes into account the generation of a suitable REST interface to access the data.

\section{\uppercase{Approach}}
\label{sec:approach}

\begin{figure*}[!ht]
	\centering
	\includegraphics[width=\textwidth]{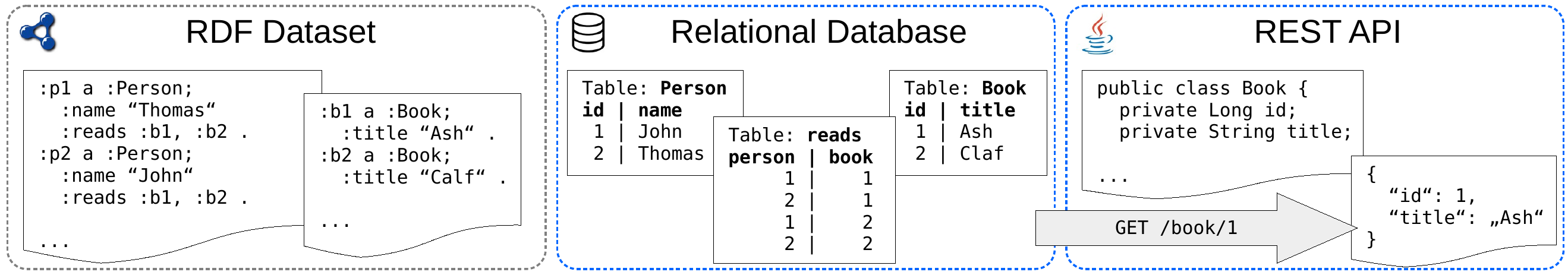}
	\caption{
		Illustrating example of our approach: An RDF dataset about persons reading books (left) is converted to a relational database (middle).
		The further generated REST API contains corresponding source code (right) and allows to query the database with REST calls (indicated by the gray arrow). The response is a JSON representation of the data.
	}
	\label{fig:approach}
\end{figure*}

\noindent Our approach is divided into three phases.
First, RDF data is analyzed to receive valuable insights about the nature of the dataset.
Secondly, these findings are used to design a suitable database model which consists of tables, columns and data records.
Thirdly, based on the database model, Java source code is generated to directly provide developers with usable data classes, a database access layer, a server and its REST interface.

\autoref{fig:approach} presents a very small example of the approach.
A simple RDF dataset about persons reading books (on the left in RDF Turtle syntax) is transformed into three tables (in the middle): 
two of them represent persons resp. books while the third one models the many-to-many relation between them. 
On the right side, corresponding source code is generated to automatically implement the REST API.
The ready to use server application can be utilized by developers to query a book resource which returns a JSON representation of it.

\subsection{Analysis of RDF}

Before the RDF model is analyzed, possible Blank Nodes in the dataset are skolemized \cite{DBLP:conf/semweb/MalleaAHP11}.
This means that they are replaced with randomly generated URIs in a consistent manner.
That way further analysis and processing is simplified without any semantic change to the RDF model.

Next, for generating the RDB tables and filling them with records later, classes and their instances have to be discovered.
They are collected by scanning through the assertion box (A-Box) statements.
In this process, instances having more than one type (multi-typed instances) are detected.
The instances' properties together with their domains and ranges are further examined since they will be modeled as either table columns or many-to-many tables. 
We divide the properties into object properties (objects are resources) and data type properties (objects are literals).
In case of object properties, the object's type is inspected (range).
If the object is not further mentioned in the RDF dataset or has no type, it is classified as a dangling resource.
Later, these resources will be stored in the database as textual URIs (instead of numeric IDs) because they are not present in the database as a record.
At least, this allows to look up such resources by following the links.
Regarding data type properties, a suitable SQL storage class is inferred which can be either text, real, integer or binary large object.
If the literal has a language tag, the property is assigned to be a special language string property.

After that, the properties cardinalities are analyzed to decide if a relationship should be modeled as a foreign key column or a many-to-many table.
For each property, based on the given data, it is deduced if it has a one-to-one, one-to-many, many-to-one or many-to-many cardinality by scanning through the A-Box statements.
The special \verb|rdf:type| property is always assumed to be many-to-many because multiple resources can have multiple types in RDF.
Additionally, all the properties' domains and ranges are collected.
Note that one predicate can have subjects and objects with various domains and ranges.
To retrieve clear mappings from domains to ranges, distinct domain-range pairs are calculated.

\subsection{Conversion to RDB}

Our generated RDBs follow the type-store approach mentioned in \cite{DBLP:journals/ker/MaCY16}.
Multi-valued attributes are avoided by using many-to-many tables when cardinalities require it.
We assume that the type-based structure is easier to grasp for developers than a horizontal or vertical structure.

Complying with the type-store approach, for each determined type from the previous step, a table is designated that will contain all instances of this type in form of records.
We denote such relations as entity tables.
They contain mandatory numeric \textit{id}-columns which serve as primary keys.
We do not add a \textit{uri}-column since the numeric primary key already serves as an identifier and databases usually do not model another textual identifier, like a URI, for their records.

An entity table is annotated with its representing RDF class. 
All properties matching the class with their domain are assumed to be a column of this table.
However, there are two special cases with respect to the cardinality of the properties.
In case of a one-to-many cardinality, the column is placed in the referring table instead.
This is a usual step when entity-relationship models (ER-models) are instantiated as relations that should satisfy the third normal form \cite{DBLP:journals/cacm/Kent83}.
In case of a many-to-many cardinality, no column in an entity table is created.
Instead, an extra table is modeled that contains two columns, namely to refer to subject and object.  
If we have a language string property at hand, another \textit{lang} column is added to store the language tag.

Next, tables are filled with data records.
To do that consistently, each resource in the RDF dataset is assigned to a unique numeric ID.
A special \textit{\_res\_id}-relation records for each URI the mapped ID for later lookup.
Using this information, records of entity tables obtain distinct IDs for their primary keys.
By scanning through the outgoing edges of every resource, record fields are allocated with the acquired objects.
For many-to-many tables, all statements with the corresponding predicate are considered.
Those statements have the following mandatory condition: 
the subject's type matches the given domain and the object's type matches the given range.
Only those statements are inserted as appropriate records in the table.
Last, SQL statements are formulated to populate an SQLite\footnote{\url{https://www.sqlite.org/}} database with the modeled tables and their records.

\subsection{Generation of REST API Code}

To generate source code and file contents, we utilize the template engine Apache FreeMarker\footnote{\url{https://freemarker.apache.org/}}.
In this phase, two Apache Maven\footnote{\url{https://maven.apache.org/}} projects are produced by our approach: 
an \textit{api} project containing data classes together with the database access layer and a \textit{server} project with the RESTful communication logic.

For the \textit{api} project, every entity table from the previous part is converted into a Plain Old Java Object (POJO)\footnote{\url{https://www.martinfowler.com/bliki/POJO.html}}.
The tables' columns are turned into attributes of the POJO class.
All many-to-many tables that could be joined with the entity table on the first column become attributes of type \verb|java.util.List|.
In case of a table that originated from a language string property, a special \verb|LangString| POJO class holding string and language tag is used as the attribute's type.

After the data classes, the database controller class is generated.
For each Java class, a corresponding \textit{select}, \textit{insert}, \textit{update} and \textit{delete} method is generated.
These methods internally use SQL to communicate with the database.
In case of the \textit{select} method, filtering is supported by using the resource query language (RQL).
For that we parse the syntax\footnote{\url{https://github.com/jirutka/rsql-parser}} and filter results based on given RQL expression.

The \textit{server} project has a dependency to the \textit{api} project. 
Thus, it reuses the POJO classes, database access logic and provides the RESTful communication.
We use the Spark\footnote{\url{http://sparkjava.com/}} framework to implement the server application.
For each Java class (from the API) a corresponding REST endpoint is generated.
The endpoints are able to interpret and perform HTTP GET, POST, PUT, PATCH and DELETE requests by utilizing the database controller from the \textit{api} project.
A converter from POJO to JSON (and vice versa) is provided to exchange data.

\section{\uppercase{Evaluation}}
\label{sec:eval}

\noindent The evaluation of our approach consists of three parts.
First, several diverse and publicly available\footnote{Except of one private dataset that was obtained from an industrial scenario.} RDF(S) datasets are transformed into RDBs to check the algorithm's ability to handle different datasets and to analyze the generated databases.
Second, we make a comparison between the outcome of a similar tool with our results using the same input dataset.
Third, a SPARQL benchmark is used which provides predefined queries and an RDF dataset generator.
Our re-engineered database is compared with the benchmark's generated SQL database.
We also investigate how well our REST API can reproduce given SPARQL queries.
The evaluation section is closed with a general discussion of the results.

\subsection{Datasets}

The Linked Open Data (LOD) cloud\footnote{\url{https://lod-cloud.net/}} is a hub that refers to all kinds of publicly available RDF-based resources or endpoints.
To test our algorithm, we randomly selected six rather small RDF datasets from the LOD cloud.
Additionally, a private dataset from an industrial scenario and a generated one from the Berlin SPARQL Benchmark (BSBM) \cite{DBLP:books/igi/11/BizerS11} are examined too.
Only relatively small sized datasets with triples around a 5-digit number of statements are chosen since we are not interested in testing the time and memory performance of our approach.
Instead, we investigate what effects various datasets have on the output of our algorithm.
The findings will be discussed at the end of the evaluation section.
Characteristics of the eight datasets and the resulting RDBs are presented in \autoref{tbl:datasets}.
In the following, we briefly describe each dataset and discuss the results individually.
\begin{table*}[t]
	\centering
	\caption{
Characteristics of eight RDF datasets and their generated databases by our approach: 
		the number of RDF statements (Stmts), classes (Cls), multi-typed instances (MT) and average number of types per MT instance (avgMT). 
		Further, the number of object property (OP), datatype properties (DP), and properties which have the following cardinality: one-to-one (OO), many-to-one (MO), one-to-many (OM) and many-to-many (MM). 
		Regarding the database, we show the number of generated entity tables (ET), many-to-many tables (MMT) and average number of columns per entity table (avgCol).
	}
\resizebox{\textwidth}{!}{\begin{tabular}{|c|l||r|r|r|r||r|r|r|r|r|r||r|r|r|}
		\hline
		No. & Name            & Stmts   & Cls   & MT    & avgMT            & OP & DP & OO & MO & OM & MM & ET   & MMT & avgCol \\
		\hline
		\hline
		1 & TBL-C             & 109     & 5   &  1    & 2                & 26 & 18 & 31 & 19   & 7 & 1  & 5   & 11       & $12.2\pm12.4$ \\\hline
		2 & CTB               & 10,853   & 4   &  0    & -                & 10 & 5  &  4 & 6    & 1 & 4  & 4   & 7        & $3.0\pm1.6$ \\\hline
		3 & EAT               & 1,674,376 & 2   &  0    & -                &  3 & 3  &  1 & 5    & 0 & 0  & 3   & 1        & $2.7\pm2.1$ \\\hline
		4 & Pokedex           & 26,562   & 19  &  0    & -                &  9 & 29 & 13 & 28   & 5 & 3  & 19  & 40       & $3.5\pm6.7$ \\\hline
		5 & BOW               & 4,041,676 & 15  & 349,195 & $ 2.9\pm0.9$    &  7 & 19 & 2  & 9    & 0 & 15 & 15  & 180      & $5.3\pm3.0$ \\\hline
		6 & S-IT              & 4,477    & 406 &  81   & $14.3\pm9.0$     &  9 & 25 & 18 & 6    & 3 & 7  & 406 & 3,056     & $2.6\pm1.0$ \\\hline
		7 & IndScn                & 25,016   & 16  &  0    & -                & 24 & 37 &  4 & 34   & 0 & 23 & 16  & 74       & $5.7\pm4.5$ \\\hline
		8 & BSBM              & 40,177   & 22  & 100   & $2.0\pm0.0$          & 12 & 28 & 16 & 22   & 0 & 2  & 22  & 17       & $13.7\pm5.5$ \\\hline
	\end{tabular}
	}
\label{tbl:datasets}
\end{table*}

\textbf{TBL-C}\footnote{\url{http://www.w3.org/People/Berners-Lee/card.rdf}} is the RDF representation of Tim Berners-Lee's electronic business card.
Since the instance representing himself has two types (\verb|foaf:Person| and \verb|con:Male|), this record is redundantly stored in two tables with identical columns.
Having multiple types causes also the generation of several many-to-many tables with equivalent data.
The reason is that for each domain-range pair a corresponding table is generated.
This effect occurs in other datasets multi-typed instances too.

The Copyright Term Bank \textbf{CTB}\footnote{\url{https://lod-cloud.net/dataset/copyrighttermbank}} dataset contains copyright terminology.
After the conversion, four entity tables contain data records: \textit{concept}, \textit{lexical\_entry}, \textit{lexical\_sense} and \textit{sense\_definition}.
However, \textit{lexical\_sense} does not have any functional properties, thus containing no columns (except the mandatory \textit{id} column).
Its records are used in a many-to-many table to group concepts by their senses.

The Edinburgh Associative Thesaurus RDF dataset \textbf{EAT}\footnote{\url{https://lod-cloud.net/dataset/associations}} \cite{DBLP:conf/esws/HeesBFBD16} contains associations of terms.
Despite the large number of statements (1,674,376), it only has two classes which results in two tables, namely \textit{association} with 325,588 records and \textit{term} with 23,218 rows.
While the \textit{association}-table has columns about \textit{stimulus} and \textit{response} together with their \textit{count} and \textit{frequency}, the \textit{term}-table has no columns (except the mandatory \textit{id} column).
This is because terms only have language string labels which are listed in a separate many-to-many table.

\textbf{Pokedex}\footnote{\url{https://lod-cloud.net/dataset/data-incubator-pokedex}} is an RDF catalog of fictitious monsters of the popular Pokémon franchise\footnote{\url{https://www.pokemon.com/}}.
The main table \textit{pokemon} contains all functional properties and has 493 entries.
All Pokémon species are listed in the \textit{species} table.
Because there is a one-to-many relationship between species (one) and Pokémon (many), the property \textit{pkm:speciesOf} is represented with a column in the \textit{pokemon} table. 
Since there is for every Pokémon type an RDF class with one instance having a label, 17 corresponding tables with only one record are created.
The consequence is that there are 34 many-to-many tables that correspond to the property \textit{pkm:type} and its inverse \textit{pkm:typeOf}.

Betweenourworlds \textbf{BOW}\footnote{\url{https://betweenourworlds.org/} (Release 2020-06)} is a dataset about animes which are drawn animations originated from Japan.
Besides having the largest number of statements in our selection (4,041,676), it also has the largest number of instances with multiple types (349,195).
The main reason is that every anime is typed with at least \textit{dbo:Anime}, \textit{dbo:Cartoon} and \textit{dbo:Work}.
As a consequence, the corresponding tables \textit{anime}, \textit{cartoon} and \textit{work} contain the exact same data records.
This causes also a lot of redundant many-to-many tables.

\textbf{S-IT}\footnote{\url{https://lod-cloud.net/dataset/salzburgerland-com-it}} is a dataset generated from the traveling website about the Salzburg state in Austria in Italian language. 
The dataset has been built with WordLift\footnote{\url{https://wordlift.io/}} \cite{DBLP:conf/esws/VolpiniR15}, which is a plugin that annotates website content with linked data.
Although, it has the second lowest number of statements (4,477), the generated database contains the most tables.
This can be explained by the high number of classes (406) and resources (81) which are instances of multiple classes.
Since these instances have in average $14.3$ types and a type corresponds to a table, they are redundantly distributed in many relations.

Our private RDF dataset \textbf{IndScn} from an industrial scenario consists of meta-data about documents and their revisions.
The conversion went straight forward: 16 classes result in the desired 16 tables with appropriate many-to-many tables for the properties.

The \textbf{BSBM}\footnote{\url{http://wifo5-03.informatik.uni-mannheim.de/bizer/berlinsparqlbenchmark/}} \cite{DBLP:books/igi/11/BizerS11} dataset has been generated with the provided generator tool (version 0.2).
Its domain is about produced, offered and reviewed products. 
With a product count parameter of 100, 40,177 RDF statements have been generated.
After the conversion, all main entity tables were created, namely \textit{product}, \textit{product\_feature}, \textit{product\_type}, \textit{person}, \textit{producer}, \textit{offer}, \textit{review} and \textit{vendor}.
Each of the 100 products has two types: the \verb|bsbm:Product| class and a certain product type instance.
That is why the number of multi-typed instances (MT) is 100 and the average number of multi-typed instances (avgMT) is exactly two.
It also causes the generation of a table per product type and a redundant storage of products.
As the review text is a language string, a many-to-many table \textit{mn\_review\_text} with a \textit{lang} (language tag) column is generated as well.

\subsection{Comparison with RDF2RDB}

RDF2RDB\footnote{\url{https://github.com/michaelbrunnbauer/rdf2rdb}} is an open-source tool that converts RDF to relational databases.
The procedure, which is written in Python 2, reads RDF files and fills a MySQL database with tables and records. 
We use two datasets, namely No. 1 and No. 8 in \autoref{tbl:datasets}, to compare exemplarily the behavior of the procedures. 

For demonstration purpose, the developer provides the generated database from Tim Berners-Lee's electronic business card\footnote{\url{https://www.netestate.de/Download/RDF2RDB/timbl.txt}}.
Since we did also the conversion of the same dataset (No. 1 in \autoref{tbl:datasets}), we can compare the resulting databases:
RDF2RDB produced 59 tables, while our procedure made only 18 relations.
The main reason for this is that RDF2RDB converts more properties into many-to-many tables.
It also generated a \textit{thing} table that sparsely records all \verb|owl:Thing| resources with their properties.
Another difference is that RDF2RDB provides a \textit{labels}-table and \textit{uris}-table.
In the \textit{labels}-relation, all URIs from the dataset are related to their labels to enable label-based searches.
The \textit{uris}-table lists for each resource its class and assigned numeric ID.
This table is comparable with our \textit{\_res\_id}-relation.

We also use the generated BSBM dataset (No. 8 in \autoref{tbl:datasets}) to compare the tool's outcome with our result. 
While RDF2RDB's database contains 145 tables, our method generated 41 relations.
RDF2RDB created for each product type a table but used for its name the product type's label.
As before, a lot of many-to-many tables are created by the tool.
The entity tables are nearly equal, except that RDF2RDB decided to model some relations as tables instead of columns.

In conclusion, RDF2RDB comply more with the RDF model with the side effect of generating more tables, especially many-to-many tables.
Our version is more data-driven, thus more properties are modeled as columns since their observed cardinalities allow that.
With this, we still do not miss any critical information in our databases.

\subsection{REST Interface Test with BSBM}

The Berlin SPARQL Benchmark (BSBM) \cite{DBLP:books/igi/11/BizerS11} provides a dataset generator and SPARQL queries to enable performance comparisons of storage systems with a SPARQL endpoint.
First, using the dataset generator, we will examine how well our approach re-engineers the benchmark's intended database.
Second, we utilize provided queries to check if our API can reproduce them.

BSBM's dataset generator can produce, aside from usual RDF, a SQL description of a database containing equivalent data.
In the following, we examine how close our method can re-engineer this database only from the RDF dataset.
Our generated database was already described in the previous section (No. 8 in \autoref{tbl:datasets}).
The comparison shows that we found all necessary entity and many-to-many tables.
Regarding the \textit{offer} table, our version misses the \textit{producer} column. 
However, this is not a surprise because the dataset does not contain any linkage between offer and producer.
Concerning the \textit{product\_type}-table, BSBM additionally added a \textit{parent} and \textit{sub\_class\_of}-column to model the class hierarchy.
Another difference is the way dates are stored.
While BSBM uses text representations in ISO 8601, our database uses milliseconds since the UNIX epoch.
Because of multi-typed instances, our procedure created some unnecessary tables per type as well as some many-to-many tables.
In conclusion, despite small differences, our re-engineered database is similarly modeled and contains all intended data.

The benchmark's main purpose is to provide queries to test performance of SPARQL endpoints.
With 12 formulated exploration queries\footnote{\url{http://wifo5-03.informatik.uni-mannheim.de/bizer/berlinsparqlbenchmark/spec/ExploreUseCase/index.html}}, endpoints can be queried in various ways by using mainly SELECT queries, a DESCRIBE query and a CONSTRUCT query.
Since they contain substitution parameters, they are actually templates that can be instantiated in various ways.  
In our evaluation, we investigate if our generated REST API can reproduce these queries by testing them with meaningful substitutions. Our expectation is that the REST API can retrieve equal information.

In the following, we present for each BSBM query its REST API call counterpart.
In some cases more then one call has to be made to join data in the client appropriately.
We omitted calls that would have been necessary to retrieve further information about referred resources, like mostly their labels.
Note that line breaks were added to fit to the text's width.

\begin{small}
\begin{alltt}
(Query 1) /product?rql=type=in=\%ProductType\%;
productFeatureProductFeature=in=
(\%ProductFeature1\%,\%ProductFeature2\%);
productPropertyNumeric1>\%x\%
(Query 2) /product/\%ProductXYZ\%
(Query 3) /product?rql=type=in=
(\%ProductType\%);
productFeatureProductFeature=in=
(\%ProductFeature1\%);
productFeatureProductFeature=out=
(\%ProductFeature2\%);
productPropertyNumeric1>\%x\%;
productPropertyNumeric3<\%y\%
(Query 4) /product?rql=type=in=(\%ProductType\%);
productFeatureProductFeature=in=
(\%ProductFeature1\%);
(productFeatureProductFeature=in=
(\%ProductFeature2\%),
productFeatureProductFeature=in=
(\%ProductFeature3\%));
productPropertyNumeric1>\%x\%;
productPropertyNumeric2>\%y\%
(Query 5a) /product/\%ProductXYZ\%
(Query 5b) /product?rql=id=out=(\%ProductXYZ\%)
(Query 6)  /product?rql=label=regex=\%word1\%
(Query 7a) /product/\%ProductXYZ\%
(Query 7b) /offer?rql=product==\%ProductXYZ\%;
validTo>\%currentDate\%
(Query 7c) /review?rql=reviewFor==\%ProductXYZ\%
(Query 8)  /review?rql=reviewFor==\%ProductXYZ\%;
text=lang=en
(Query 9)  /review/\%ReviewXYZ\% 
(Query 10) /offer?rql=product==\%ProductXYZ\%;
deliveryDays<=3;validTo>\%currentDate\%
(Query 11) /offer/\%OfferXYZ\%
(Query 12) /offer/\%OfferXYZ\%
\end{alltt}
\end{small}
The resource query language (RQL) is often used to mimic SPARQL's basic graph patterns and filter possibilities.
We put the responsibility for ordering the results to the client.
The interpretation of optional information (expressed in SPARQL with the OPTIONAL keyword) is also up to the developer.

What follows are short explanations how some queries are reproduced with our API.
Regarding Query 3, the label of the optional \verb|ProductFeature2| is checked to be not bound.
This is solved in our case by excluding this feature with RQL's \verb|=out=| operator which yields to the same result.
In Query 4, the UNION statement is imitated by a logical \textit{or} construction (in RQL a `\verb|,|') whether \verb|ProductFeature2| or \verb|ProductFeature3| (or both) are the product's features.
Concerning Query 5, the SPARQL query contains filters with arithmetic expressions that can not be emulated by our RQL engine.
Thus, in a second call (5b), the client will be responsible for filtering the results to find matching products.
Using RQL's \verb|=regex=| operator, we reproduce the regular expression filter in Query 6.
Query 7 demonstrates that joining data involves multiple REST calls and has to be done on the client-side.
Since data types of literals are not stored in our database, the currency of an offer's price can not be retrieved.
With the special \verb|=lang=| operator in RQL, language strings can be filtered as shown in Query 8.
Regarding Query 9, a first call determines the reviewer of \verb|ReviewXYZ|, while in a second call (not shown in the code), the reviewer is retrieved by using the \verb|/person| endpoint.
In Query 11, also incoming edges of a given \verb|OfferXYZ| are queried.
Since our REST API can only retrieve outgoing edges of a resource, this request is only partially reproducible.
In case of Query 12, which is a CONSTRUCT query, the actual construction part lies in the responsibility of the client.

\subsection{Discussion}

At the beginning of the paper, we stated three question that can now be answered based on the evaluation results.
When answering the first two questions, we also address limitations of our generated databases and interfaces.

\textit{How well do the generated RDBs reflect their RDF dataset counterparts?}
The observation of the eight generated databases (\autoref{tbl:datasets}) shows that the information content of a dataset's A-Box is sufficiently reflected by its database counterpart, i.e. we did not miss any critical information.
However, it is possible to express the same information with different relational models. Compared to RDF2RDB, our procedure created less tables but still more than intended by BSBM.
The evaluation reveals that the handling of multi-typed instances poses the main challenge.
Since RDF classes directly correspond to RDB tables, a resource having more then one type is redundantly distributed among respective tables.
It also causes the generation of more many-to-many tables because such a table relates one certain domain to one particular range.
That means that unnecessary and unwanted data redundancies occur.

Another major challenge is the decision for a trade-off that determines whether a property becomes a table or a column.
If, on the one extreme, every property has a table counterpart, the number of tables explodes and joins become inevitable which make queries more complex.
If, on the other extreme, properties become columns in tables, data redundancy occurs since relations would violate the second normal form. That is why generation procedures should have a meaningful (maybe configurable) trade-off.
We decided to infer the cardinality of properties based on existing A-Box statements to minimize the number of many-to-many tables.
However, this makes the properties' cardinalities unchangeable in the database model.
For example, a one-to-many relationship can not be easily turned into many-to-many relationship once the database model is defined.

Evaluation also points out smaller issues in our generator.
Tables having only an \textit{id} column could be removed because they provide no further information.
We also noticed that RDF data types are missing in our database model.
In conclusion, our generated RDBs indeed reflect their RDF dataset counterparts sufficiently but contain unnecessary redundant data as well.

\textit{How well can the generated CRUD REST API reproduce queries that would have been performed with SPARQL?}
Many SPARQL operations like aggregates, arithmetic expressions, sub-queries and various functions are not supported by our rather simple API.
When such queries become more complex, we make the client responsible to send more requests and to process the results accordingly. A major issue is that the API does not provide a join-mechanism.
Since each endpoint represents a class and returns the corresponding instances, the join has to be performed by the client.
Another limitation is the inability to retrieve a resource's incoming edges since the underlying database model is not designed for that.
Yet, by using the BSBM benchmark, we showed that our interface could in almost all cases retrieve the same information as the given SPARQL queries.
Hence, our conclusion is that the API can reproduce rather simple and common queries, while more complex ones have to be handled by the client.

\section{\uppercase{Conclusion and Outlook}}
\label{sec:concl}

\noindent Especially in industry, Semantic Web technologies are rather seldom used. 
Since corporations would greatly benefit from available and future RDF-based datasets, we suggested to bridge the technology gap by a fully automatic conversion of RDF to RDBs together with CRUD REST APIs.
Our approach to accomplish that consists of three steps: the analysis of a given RDF dataset, the conversion to a relational model and the generation of source code to implement a RESTful server logic.
We conducted several experiments. 
In comparison to related work, our re-engineered databases reflect their RDF counterparts with less tables.
Moreover, our generated REST APIs are able to reproduce rather simple and common SPARQL queries.
We identified as a remaining challenge the generation of unnecessary data redundancies because of multi-typed instances.

Future work should find an appropriate way to model the database to reduce the high number of tables and data redundancies. In this regard, as already pointed out, algorithms should provide a configurable trade-off to decide whether properties become many-to-many tables or simple columns.
Thus, the right setting can be dependent on a particular use case. 
Since RDF datasets can change over time, we also suggest that future procedures provide an update-mechanism in order to avoid rebuilding the whole database (and possibly removing already inserted data).
Moreover, to also process larger datasets (like DBpedia), we intend to reduce our algorithm's memory usage.

\bibliographystyle{apalike}
{\small \bibliography{paper}}

\end{document}